# Development of a multi-pixel hybrid photo-detector with high quantum efficiency and gain

M. Suyama, A. Fukasawa, J. Haba, T. Iijima, S. Iwata, M. Sakuda, T. Sumiyoshi, F. Takasaki, M. Tanaka, T. Tsuboyama and Y. Yamada, *non-Member, IEEE*

*Abstract*— A hybrid photo-detector (HPD) consisting of a photocathode and a multi-pixel avalanche diode (MP-AD) was developed a few years ago. Our previous studies showed that its inherent potential for high resolution photon counting could be further enhanced by reducing fluctuations in charge loss in the dead layer at the entrance of the MP-AD. In this paper, we report on the improvement with the newly developed HPD whose encapsulated MP-AD has a thinner dead layer than before. It is demonstrated that the new HPD has much better energy resolution, which enables clearer counting up to nine photoelectrons. Further enhancement of the photocathode sensitivity of the HPD is also discussed.

## I. INTRODUCTION

MULTI-PIXEL photon sensors with single-photon sensitivity are widely used in high-energy physics detectors, such as scintillating fibers [1] or Cherenkov counter readout [2], [3]. Broader applications are expected if the sensors can operate in a strong magnetic field [4], [5]. Such photon sensors can be realized with hybrid photo-detector (HPD) technology consisting of a photocathode and a multi-pixel avalanche diode (MP-AD). We succeeded in building an HPD having 8x8 pixels so far [6], [7], with a gain as high as $5 \times 10^4$, sufficient to detect a single photon with a timing resolution of better than 100 ps. Operation in a magnetic field up to 1.5 T was confirmed as well.

In previous studies [6], one of the important suggestions is that the dead layer at the entrance of the MP-AD plays a crucial role for the energy resolution. We now demonstrate improved performance of the HPD with a new MP-AD having a thinner dead layer compared to the one (old MP-AD) developed in the previous study.

## II. STRUCTURE OF DEVELOPED HPD

A schematic drawing of the newly developed HPD is shown in Fig. 1. The primary components of the HPD are a photocathode and an MP-AD. The multi-alkali photocathode is formed on the vacuum side of the input window. The MP-AD is assembled on the ceramic stem facing the photocathode with a small distance of 2.5 mm to minimize the effect of external magnetic fields. This proximity focusing structure is essential for vacuum tubes to be used in a high magnetic field. Output pins come out on the back of the stem where a bias voltage to the MP-AD can be applied or signals can be extracted. A photograph of the tube is shown in Fig. 2.

The MP-AD has a reach-through structure whose backside is irradiated by electrons, as shown in Fig. 3. The p+ layer is formed on a Si substrate, where the density of p type impurity, boron, is high enough to make an ohmic contact. An $8 \times 8$ array of independent pn junctions, each of which forms a pixel, are fabricated on the opposite surface of the p+ layer. With careful adjustment of the impurity density of the p layer, the electric field around the pn junction is high enough to multiply the collected electrons by the avalanche process. The size of one pixel is 2 mm $\times$ 2 mm, and the total area is 16 mm $\times$ 16 mm. An advantage of the reach-through structure with back irradiation is that the p+ layer at the entrance of electrons can be thinned without practical difficulties, because it is independent of the pn junctions governing the delicate avalanche multiplication.

In response to incident photons, electrons are emitted from the photocathode, accelerated and hit the MP-AD. Each of the electrons gives rise to ionization there and produces an electron-hole pair for every 3.6eV of the deposited energy, amounting to thousands of pairs. This multiplication is referred

Manuscript received October 29, 2003. This work was supported in part by Grant-in-Aid for Scientific Research (B)(2) 13554008 of Japanese Ministry of Education, Culture, Sports, Science and Technology.

M. Suyama and A. Fukasawa are with Electron Tube Center, Hamamatsu Photonics K.K., Shimokanzo 314-5, Toyooka-village, Iwata-gun 438-0193, Japan (telephone: 81-539-62-3151, e-mail: suyama@etd.hpk.co.jp, fukasawa@etd.hpk.co.jp).

J. Haba, S. Iwata, M. Sakuda, F. Takasaki, M. Tanaka, T. Tsuboyama and Y. Yamada are with Institute of Particle and Nuclear Studies, High Energy Accelerator Research Organization (KEK), Oho 1-1, Tsukuba 305-0801, Japan (telephone: 81-29-864-1171, e-mail: junji.haba@kek.jp, seigi.iwata@kek.jp, makoto.sakuda@kek.jp, fumihiko.takasaki@kek.jp, manobu.tanaka@kek.jp, toru.tsuboyama@kek.jp, yoshikazu.yamada@kek.jp).

T. Iijima is with Department of Physics, Nagoya University, Furou-cho, Chikusa-ku, Nagoya 464-8602, Japan (telephone: 81-52-789-2893, e-mail: iijima@hepl.phys.nagoya-u.ac.jp).

T. Sumiyoshi is with Department of Physics, Tokyo Metropolitan University, Minami-Ohsawa 1-1, Hachiouji-shi, Tokyo 192-0364, Japan (telephone: 81-426-77-2514, e-mail: sumiyoshi@phys.metro-u.ac.jp).

to as electron-bombarded gain. The generated electrons then drift along the electric field inside the MP-AD toward individual pn junctions, where each electron induces avalanche multiplication, which is read out though the output pins.

The previous study [7] suggests that p+ layer at the entrance forms a dead layer and the energy deposited there does not contribute to the electron-bombarded gain. Since the fluctuation of deposited energy in the dead layer predetermines the overall energy resolution, the thinnest possible dead layer is desirable. In this study, a thinner dead layer than before[6] is realized by low energy implantation of a p-type impurity, boron.

### III. TEST RESULTS

#### A. Electron-bombarded gain

The electron-bombarded gain was evaluated as a ratio of the output current from the MP-AD to the photocurrent from the photocathode at an avalanche gain of unity. The electron-bombarded gains of HPDs with the new MP-AD (black circles) and the old one (open triangles) are shown in Fig. 4, as a function of photocathode voltage. The electron-bombarded gain is improved from 1600 for the old MP-AD to 2100 for the new type at a photocathode voltage of –9kV.

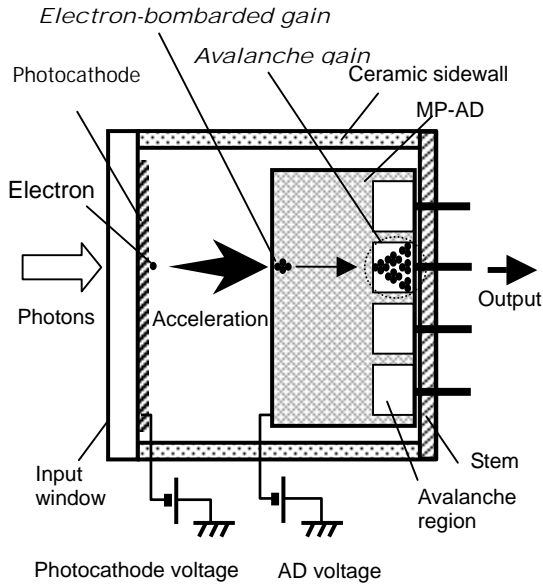

Fig. 1. The developed HPD is schematically shown. Primary components are the photocathode and the MP-AD. 8kV high voltage is applied between them to obtain electron-bombarded gain.

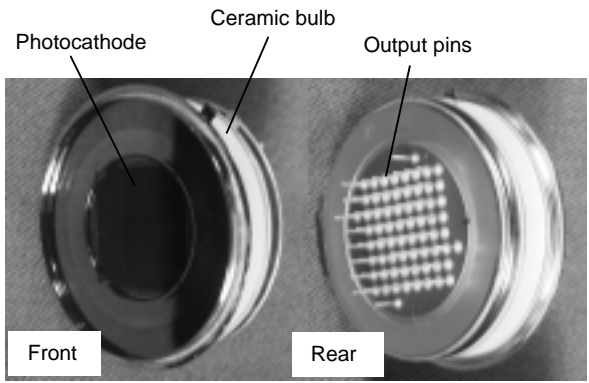

Fig. 2. A photograph of the developed HPD is shown.

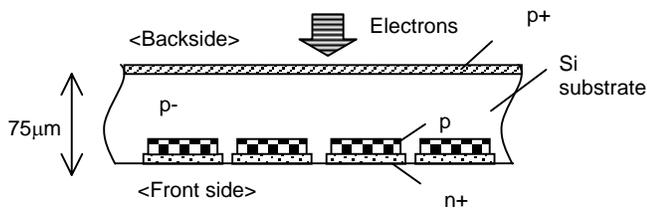

Fig. 3. A schematic drawing of the MP-AD encapsulated in the HPD is shown. Electrons bombard the MP-AD from the p+ side and generate electron-hole pairs. The generated electrons drift toward pn junctions, where the electrons are further multiplied by avalanche multiplication.

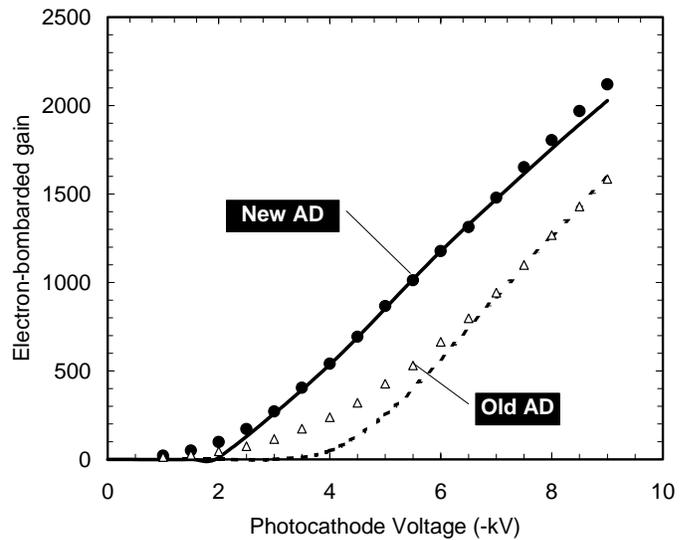

Fig. 4. Electron-bombarded gains of HPDs with the new MP-AD (black circles) and the old one (open triangles) are shown as a function of the photocathode voltage. The electron-bombarded gain is improved from 1600 (old one) to 2100 (new one) at the photocathode voltage of –9kV. Gains were simulated by varying the thickness of the dead surface layer. The best fit for the new AD is obtained with a thickness of 80nm, as shown by solid line, whereas for the old one the best fit is obtained with 210nm, as shown by dotted line.

The electron-bombarded gain was fitted with the expectation from the simulation code, PENELOPE [8], which can simulate transport of low energy electrons and photons. By adjusting the thickness of the dead layer in the simulation, we found that a thickness of 80 nm gives the best description of the new MP-AD, whereas 210 nm is best for the old one, as shown in Fig. 5 with solid and dotted lines, respectively. The disagreement between data and simulation for the old type in the low voltage region may be the result of the simple assumption that no electrons generated in the dead layer could reach the avalanche region. The agreement is better for the new one since the effects here are expected to be smaller as less energy is deposited in the dead layer.

### B. Avalanche gain and leakage current of MP-AD

For the measurements of avalanche gain and leakage current, all pixels were connected in parallel to sum up the signals from the pixels. Fig. 5 shows avalanche gains of the new MP-AD (solid line) and the old one (dotted line) measured with incident photoelectrons at a photocathode voltage of –8kV. The AD gains are almost the same within the deviations expected for the variations in the impurity density of the p layer facing the n+ layer, which would significantly change the electric field in the avalanche region, and thus affect avalanche gain greatly. The maximum reachable avalanche gains of these MP-ADs are limited by the breakdown voltage of the worst pixel contained. Without a bad pixel, the avalanche gain reaches 100, as shown in [7].

Leakage currents for the new MP-AD and the old one as a function of the bias voltage are shown in Fig. 6. There is a rapid increase of the current for the new MP-AD around 200V due to a bad pixel. Below the breakdown point, the new MP-AD shows an order of magnitude higher leakage current than that of the old one. It is possible that the thinner p+ layer may not be sufficient to prevent thermal charges from flowing out as an excess dark current. Further optimization of the process is in progress.

### C. Pulse height spectrum for multi-photons

The pulse height spectra for multi-photons were measured with HPDs illuminated by a pulsed LED of 470 nm. A charge-sensitive amplifier and a linear amplifier with a shaping time of 0.5μs was used to readout the signal from the one pixel while the other pixels were connected to the ground. Both the HPD with the new MP-AD and the old one were operated with the same total gain, approximately $4.5 \times 10^4$ at a photocathode voltage of –9 kV.

The results are shown in Fig. 7, where the black curve is the data with the new MP-AD, and the gray curve with the old one. The pulse height spectrum is improved drastically for the HPD encapsulating the new MP-AD with the thinner dead layer. This improvement clearly proves our conjecture on the energy resolution of the HPD.

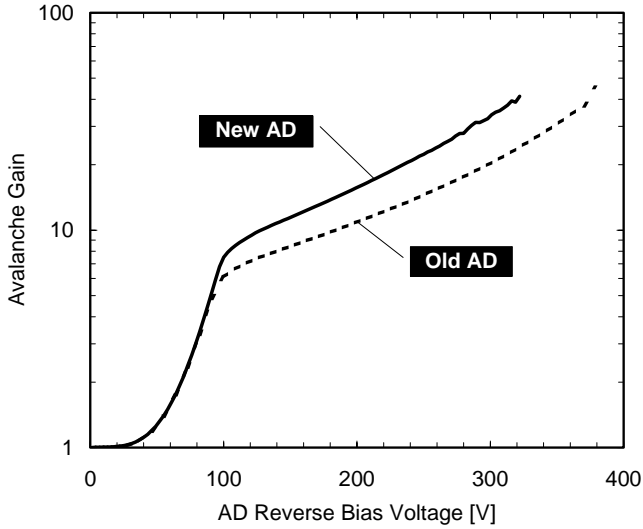

Fig. 5. Avalanche gains for the new MP-AD (solid line) and the old one (dotted line) are shown as a function of the applied bias voltage.

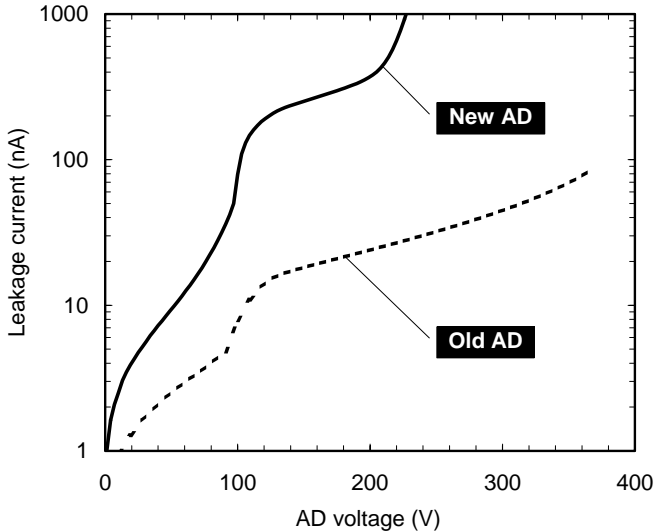

Fig. 6. Leakage current for the new MP-AD (solid line) and the old one (dotted line) are shown as a function of the applied bias voltage.

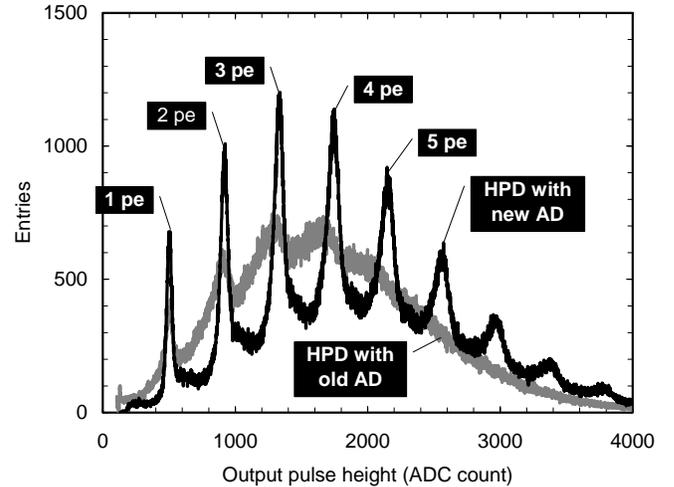

Fig. 7. Pulse height spectra of HPDs with the new MP-AD (black) and the old one (gray) for multi-photon input.

### D. Uniformity

The output current from each pixel of the HPD with the new MP-AD was measured by scanning a light spot on the center of each pixel, where the diameter of the spot was 1mm and wavelength was 500 nm. The voltages on the photocathode and the MP-AD were –8 kV and 250V, respectively, for a total gain of $4.5 \times 10^4$.

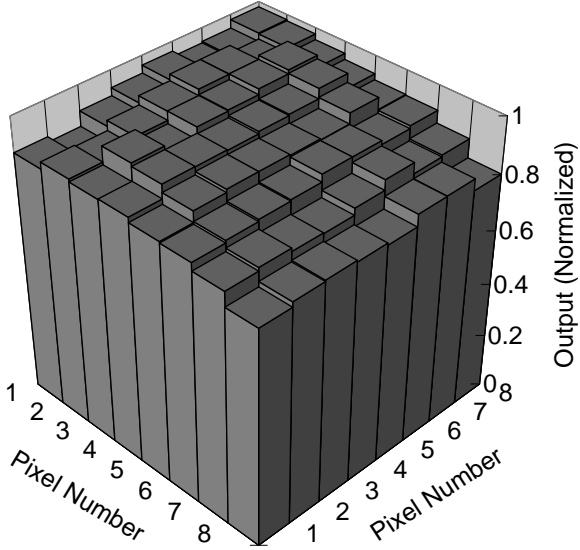

Fig. 8. Relative output of each pixel measured by scanning a spot of light over the center of each pixel is shown. The standard deviation normalized to the average output is 6.0%.

The relative output from each pixel is shown in Fig. 8. The standard deviation among them is 6.0%, when it is normalized to the average output signal. This uniformity is better than the value of 30% reported for a multi-anode PMT commercially available [3]. Sources of this non-uniformity are attributed to the photocathode sensitivity, the electron bombarded gain and the avalanche gains as reported in [6], [7].

## IV. EFFORT TO IMPROVE PHOTOCATHODE SENSITIVITY

Some attempts are being made to adapt a photocathode with a higher sensitivity made of GaAsP crystal to the HPD. Since the mean free path of electrons in the crystal is long, the photocathode can be thick enough to absorb photons without major loss of excited electrons diffusing to the vacuum side. The negative electron affinity of the GaAsP photocathode leads to high emission probability from the crystal to vacuum. The GaAsP photocathode, thus, can achieve high sensitivity for photons.

The best quantum efficiency (QE) curve measured for the preliminary test tubes with a GaAsP photocathode (solid curve) is shown in Fig. 9, together with that of multi-alkali photocathode (dashed line) for reference. The sharp cut-off of the GaAsP photocathode around 300nm is due to the transmission characteristics of the borosilicate glass used for the input window to match its thermal expansion coefficient to the GaAsP crystal, while no such cut-off is observed with quartz glass used for the multi-alkali photocathode. This demonstrates that the QE of a GaAsP photocathode can be as high as 60%. The existing prototype HPD with the same type of photocathode shows the QE of 32% for now. Optimization of the photocathode process to the HPD is still in progress.

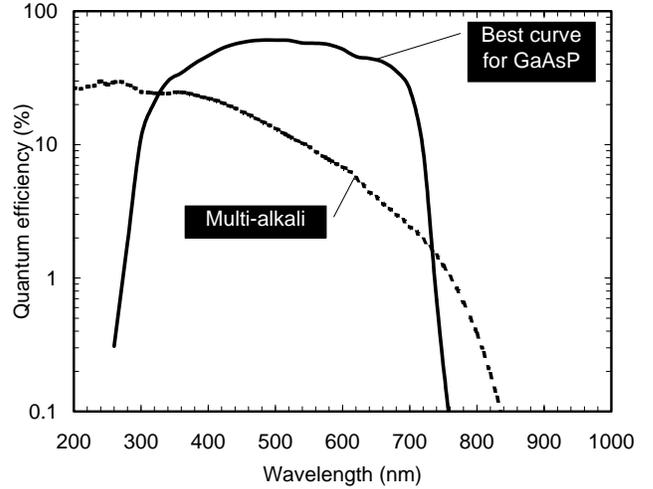

Fig. 9. The best QE curve for the tubes with the GaAsP photocathode (solid line) is shown. QE for a multi-alkali photocathode developed in the previous work [6] is shown by the dashed line for reference.

## V. SUMMARY

Performance of a newly developed HPD with an avalanche diode is compared with that encapsulating a diode [9], as listed in Table I. A remarkable feature of the new HPD is one order of magnitude higher gain than the one with the diode, even with lower photocathode voltage. This enables single photon detection with high-speed amplifiers, which have inherently large noise.

Features of a HPD with AD are further enhanced by the thinner dead layer at the entrance of the AD. Thanks to the smaller fluctuations of the energy deposit in the layer, the energy resolution is drastically improved and is now sufficient to clearly identify up to nine photoelectrons. It is also demonstrated that a high quantum efficiency (up to 60%) HPD can be made with a highly sensitive GaAsP photocathode.

TABLE I
COMPARISON OF HPDS WITH AN AD AND A DIODE [9]

|  | HPD with avalanche diode | HPD with diode [9] | Unit |
|---|---|---|---|
| Pixel size | 4 | 3.5 | $mm^2$ |
| Pixel number | 64 | 61 | pixels |
| Capacitance | 6 [6] | 4 | pF |
| Dead zone between pixels | 0.12 [6] | 0.05 | mm |
| Photocathode voltage | -9 | -14.5 | kV |
| Diode voltage | 250 | 60 | V |
| EB gain | 2,100 | 3,750 |  |
| Avalanche gain | 21 | - |  |
| Total gain | 45,000 | 3,750 |  |

## VI. Acknowledgment

We would like to express our thanks to Messrs. K. Yamamoto, Y. Ishikawa and M. Muramatsu of the solid state division of Hamamatsu Photonics K.K. (HPK) for technical discussions about avalanche diodes. We greatly appreciate the efforts of Messrs. A. Kageyama and K. Inoue of the electron tube center (ETC) of HPK for assembling the avalanche diode, and Mr. Y. Kawai of the ETC of HPK for discussions about the evaluation. We are also grateful to Prof. T. E. Browder of University of Hawaii for the improvement of our manuscript. This work was partially supported by Grant-in Aid for Scientific Research (B) (2) 13554008 of the Japanese Ministry of Education, Culture, Sports, Science and Technology.

## VII. References

[1] K. Rielagea, K. Arisakab, M. Atacb, W.R. Binnsa, M.J. Christlc, P. Dowkontt et al., "Characterization of a multianode photomultiplier tube for use with scintillating fibers", *Nucl. Instrm. Methods*, vol. A 463, pp. 149–160, 2001.

[2] T. Gys, "The pixel hybrid photon detectors for the LHCb-rich project", *Nucl. Instrm. Methods*, vol. A465, pp. 240-246, 2001.

[3] E. Albrechta, J. Bakerb, G. Barberc, J. Bibbyd, M. Calvie, M. Charles et al., "Performance of a cluster of multi-anode photomultipliers equipped with lenses for use in a prototype RICH detector", *Nucl. Instrm. Methods*, vol. A 488, pp. 110–130, 2002.

[4] P. Cushman, A. Heering, J. Nelson, C. Timmermans, S. R. Dugad, S. Katta et al. "Multi-pixel hybrid photodiode tubes for the CMS hadron calorimeter", *Nucl. Instrm. Methods*, vol. A387, pp.107-112, 1997.

[5] M. Akatsu, M. Aoki, K. Fujimoto, Y. Higashino, M. Hirose, K. Inami et al., "Time-of-Propagation Cherenkov counter for particle identification", *Nucl. Instrm. Methods*, vol. A440, pp.124-135, 2000.

[6] M. Suyama, A. Fukasawa, J. Haba, T. Iijima, S. Iwata, M. Sakuda et al., "Development of a multi-pixel photon sensor with single-photon sensitivity", *Nucl. Instrm. Methods*, , to be published.

[7] M. Suyama, "Development of a multi-pixel photon sensor with single-photon sensitivity", *KEK Report*, vol. 2002-16, pp.1-115, 2003.

[8] F. Salvat, J. M. Fernandez-Varea, E. Acosta and J. Sempau, "PENELOPE – A Code System for Monte Carlo Simulation of Electron and Photon Transport", *NEA/NSC/DOC*, Vol. 19, pp.1-***, 2001.

[9] C. Damiani, A. Del Guerra, R. De Salvo, G. Vintale, G. Di Domenico, et al., "Magnetic field effects on Hybrid PhotoDiode single electron response", *Nucl. Instrm. Methods*, vol. A422, pp.136-139, 2000.